\documentclass[%
 reprint,
superscriptaddress,
 amsmath,amssymb,
 aps,
prl,
floatfix,
]{revtex4-1}
\usepackage{amssymb}
\usepackage{mathrsfs}
\usepackage{graphicx}
\usepackage{dcolumn}
\usepackage{bm}
\usepackage{amsmath}
\usepackage{amsfonts}
\usepackage{color}
\usepackage{float}
\usepackage{graphicx}% Include figure files
\usepackage{dcolumn}% Align table columns on decimal point
\usepackage{bm}% bold math
\usepackage{hyperref}% add hypertext capabilities
\usepackage{xcolor}
\hypersetup{
	colorlinks,
	linkcolor={red!50!black},
	citecolor={green!50!black},
	urlcolor={blue!80!black}
}
\usepackage[caption=false]{subfig}
\makeatletter
\newcommand{\colorcaption}[2][]{%
  \begingroup%
  \renewcommand{\@caption@fignum@sep}{ (color online). }%
  \caption[#1]{#2}%
  \endgroup%
}
\makeatother

\begin{document}

\preprint{APS/123-QED}

\title{Non-factorizable 4D quantum Hall state from photonic crystal defects}% Force line breaks with \\

\author{Xiao Zhang}
 \email{zhangxiao@mail.sysu.edu.cn}
\affiliation{School of Physics, Sun Yat-sen University, Guangzhou 510275, China}

\author{Youjian Chen}
\affiliation{School of Physics, Sun Yat-sen University, Guangzhou 510275, China}

  \author{Bochen Guan}
% \email{guanboch@mail2.sysu.edu.cn}
\affiliation{Department of Electrical and Computer Engineering, University of Wisconsin, Madison, WI 53706, USA}

 \author{Jun Yu Lin}
%\email{guanboch@mail2.sysu.edu.cn}
\affiliation{School of Physics, Sun Yat-sen University, Guangzhou 510275, China}
\affiliation{Yat-Sen School, Sun Yat-sen University, Guangzhou 510275, China}

 \author{Nai Chao Hu}
% \email{hunch@mail2.sysu.edu.cn}
 \affiliation{School of Physics, Sun Yat-sen University, Guangzhou 510275, China}
\author{Ching Hua Lee}
% \email{hunch@mail2.sysu.edu.cn}
 \affiliation{Institute of High Performance Computing, 138632, Singapore}

\date{\today}% It is always \today, today,

\begin{abstract}
In the recent years, there has been a drive towards the realization of topological phases beyond conventional electronic materials, including phases defined in more than three dimensions. We propose a way to realize 2nd Chern number topological phases with photonic crystals simply made up of defect resonators embedded within a regular lattice of resonators. In particular, through a novel quasiperiodic spatial modulations in the defect radii, a defect lattice possessing topologically nontrivial Chern bands with non-abelian berry curvature living in four-dimensional synthetic space is proposed. This system cannot be factorized by a direct product of two 1st Chern number models, distinguishing itself from the Hofstadter model. Such photonic systems can be easily experimentally realized with regular photonic crystals consisting of dielectric rods in air.
\end{abstract}

\pacs{Valid PACS appear here}% PACS, the Physics and Astronomy
\maketitle

\section{\label{sec:level1}Introduction}Topological phases like quantum Hall systems \cite{von1986quantized,novoselov2007room,jain1989composite}, topological insulators\cite{qi2010quantum,qi2011topological,hasan2010colloquium,bernevig2006quantum,zhang2009topological,zhang2012actinide,He03052016} and quantum anomalous Hall materials \cite{chang2013experimental,yu2010quantized,liu2008quantum,ezawa2012valley} rank amongst the most intensely studied topics in condensed matter physics, material science and electrical engineering communities\cite{zhang2010topological,barsoum2000electrical}. In these systems, each energy band is ascribed a topological index that mandate the presence of interesting boundary states and quantized response. Soon after the discovery of such topological phases in electronic condensed matter systems, analogs in phononic (acoustic) \cite{prodan2009topological,kane2014topological,susstrunk2015observation,nash2015topological,yang2015topological,yang2015topological2,fleury2015floquet,khanikaev2015topologically,wang2015topological,rocklin2015transformable,liu2016topological,cheng2016robust,He2016,ong2016transport} systems, photonic systems \cite{haldane2008possible,wang2008reflection,wang2009observation,rechtsman2013photonic,gao2015topological,PhysRevLett.114.223901,WeylPhotonicLilu,WeylPhotonicLiluScience,lu2016symmetry,Xu2016,lin2016dirac} or their hybrid\cite{peano2015topological} have been proposed and in some cases realized experimentally. Although topological phases with nontrivial Chern invariant in principle exist in any even number of dimensions, physical real space is limited by three dimensions. Hence all conventional topological Chern systems in two or three dimensions, i.e. quantum anomalous Hall materials, Weyl semimetals\cite{WeylsemimetalHongMing,WeylsemimetalHasan} and their photonic analogs \cite{haldane2008possible,wang2008reflection,kraus2012topological,verbin2013observation,kraus2012topological2,WeylPhotonicLilu,WeylPhotonicLiluScience,XuefengZhu,Rechtsman} can only be characterized by the first Chern number. Topological phases characterized by the second Chern number, dubbed four dimensional (4d) quantum Hall systems\cite{zhang2001four} cannot be realized in 3d physical space, and have for a long time been believed to exist only in theory. 

Recently, it was realized that such second Chern number phases can be realized in photonic systems through synthetic dimensions. Proposals include uni-directional propagating waveguides with aperiodic structures\cite{kraus2013four} and optical fibers of Weyl materials with helical structure\cite{lu2016topological}. In the former, the synthetic dimensions arise through the identification of an aperiodic hamiltonian with a Hofstadter lattice containing twice the number of dimensions\cite{hatsugai1993edge,dean2013hofstadter,madsen2013topological,satija2014hidden,tran2015topological,fuchs2016hofstadter}. This aperiodic PC is described by an effective tight-binding Hamiltonian corresponding to the tensor product of two Hofstadter models with irrational fluxes. The resultant band structure assumes a fractal form, and can be characterized by the second Chern number in the space of synthetic dimensions. While first Chern number edge states have been experimentally observed in 1d analogs of such propagating waveguides\cite{kraus2012topological}, the experimental realization of their 2d (second Chern number) analogs remain elusive. Further more, a photonic system with a 2nd number which can not be factorized into two models with 1st Chern number remains to be discovered. As such, in this paper we propose an alternative realization of a state with an either factorizable or non-factorizable second Chern number that is based on conventional photonic crystals (PCs), which is highly desirable due to its maturity in experimental setup.

In this work, we construct topologically protected defect modes in a regular two dimensional (2d) PC by simply aperiodically modulating the resonator's sizes. To motivate our approach, we begin by illustrating the mid-gap defect modes\cite{Quasicrystal2,Quasicrystal3,Quasicrystal1,Quasicrystal4,Maser1,Maser2,Maser3,villeneuve1996microcavities,Rechtsman,PhotonicApplication1,PhotonicApplication2,PhotonicApplication3} of the Hofstadter model protected by the factorizable second Chern number, in a 2d resonator lattice embedded in a 2d PC. The protected defect modes, localized at the boundary of the 2d lattice, can be viewed as boundary modes in a 4d synthetic space. Next we generalize to a novel modulated resonator lattice in a 2d PC, and achieve mid-gap defect modes protected by the non-factorizable second Chern number with non-abelian berry curvature. Our calculations are performed by exactly solving Maxwell equations via the finite element method (FEM) with the Maxwell Equation solver COMSOL. Finally we propose simple experimental setups in the microwave wavelength window, with proposed PC structures of centimeter size, and which can be easily manufactured with modern fabrication technology.

\section{Tight-binding Hamiltonian of a 2d resonator lattice}
\begin{figure}[h]
\centering
\includegraphics[width=\linewidth]{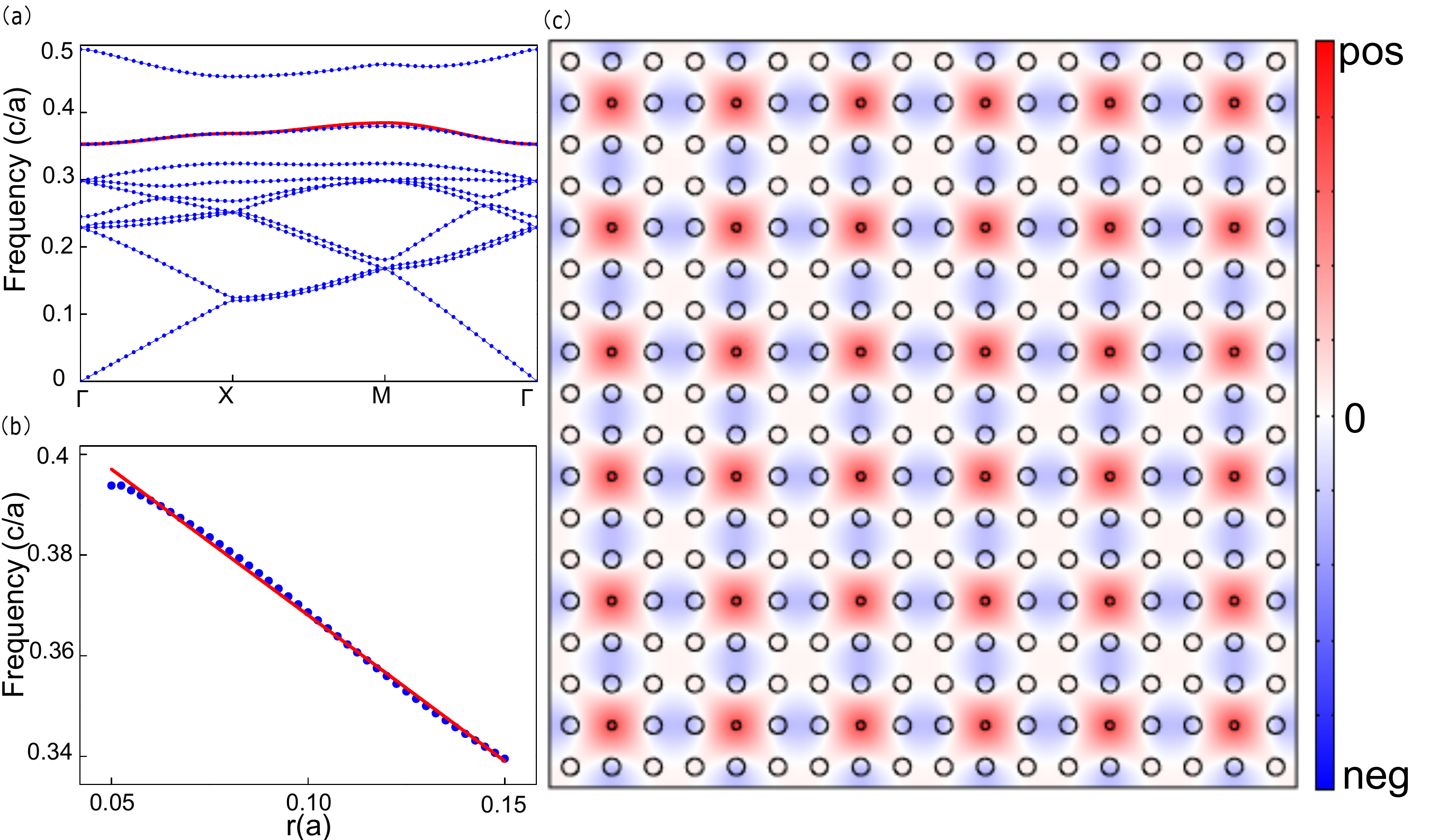}
\caption{ (a) The frequency bands for the TM modes of our PC consisting of a 2d lattice of radius $r=0.2a$ rods. Embedded within it is a linear array of defect rods with radius $r=0.1a$ that replaces one in three regular rods. The dielectric constant is 8. The defect state (red) lies in the large band gap of bulk states (blue). By fitting the energy band with simulation data we found $\omega_0=0.36862(c/a)$ and $t=-0.00398(c/a)$. (b) The monotonic dependence of the frequency of the defect state at the $X$ point and the radius of defect rods. By fitting the curve 
We found $a=-0.58032(c/a^{2})$ and $b=0.42608(c/a)$. (c) The distribution of the $z$ component of the electric field strength of the defect states at the $\Gamma$ point. It decays rapidly away from the line of defects. The maximum value in the color bar is 5V/m while the minimum value in the color bar is -5V/m.}  
 \label{fig:1}
 \end{figure}

To begin with, we give a pedagogical description of the defect modes in 2d PC using a tight-binding (TB) Hamiltonian. First we convert Maxwell's equation to a tight-binding Hamiltonian in the basis of defect modes\cite{wang2005magneto} (Fig \ref{fig:1}):
\begin{equation}
H|\psi\rangle =\begin{bmatrix}
0 & i\varepsilon_{0}^{-1}\vec\nabla\times\\
-i\mu_{0}^{-1}\vec\nabla\times & 0\\	
\end{bmatrix}\begin{bmatrix}
\vec{E}\\
\vec{H}
\end{bmatrix}=\omega|\psi\rangle 
\end{equation}
where $|\psi\rangle =\begin{bmatrix}
\vec{E}\\
\vec{H}
\end{bmatrix}$
denotes an eigenmode with eigenfrequency $\omega$. $\varepsilon_0(x,y)$ is a function of the refractive index in real space. Our PC consists of a lattice of equally spaced rods of dielectric constant $8$ and radius $0.2a$ separated by air, where $a$ is the lattice spacing. Embedded in this 2d PC is a 1d resonator chain in the $x$ direction made up of rods with radius smaller than $0.2a$, each separated by 2 regular rods (Fig. \ref{fig:1}c).

Fig. \ref{fig:1} shows the solution to the Maxwell's equation in the 2d resonator chain.  For the TM modes, the 2d PC regular rods form a gapped bulk state which can seen as the "vacuum" surrounding the near-flat defect mode appearing in the mid-gap (Fig. \ref{fig:1}a). The defect mode is a resonator confined electromagnetic mode(Fig. \ref{fig:1}c), which can be viewed as a photonic analog to an orbital wave function of an atom. Collectively they form the TB basis of the effective Hamiltonian of the PC. Thus we will be able to describe the defect modes with a TB model\cite{wang2005magneto,Hamiltonian1} consisting of the on-site energy $\omega$ at every site $i$, and hopping amplitude $t$ between nearest neighborhood sites $i,j$: 
\begin{equation}
H=\sum_{i}\omega_0|\psi_{i}\rangle \langle \psi_{i}|+\sum_{\langle i,j\rangle }t|\psi_{i}\rangle \langle \psi_{j}|
\end{equation}
Further hoppings can be neglected as the defect modes are well-localized. Using the Bloch Theorem we can obtain the energy band of the localized state
\begin{equation}
E(k_x,k_y)=\omega_0+2t(cos(2ak_x) +2t(cos(2ak_y)
\end{equation}
When changing the radius of the central rod the coupouling constant t changes slowly but the on-site energy $\omega_0$ changes dramatically. To some extent,the relation between the radius of the central rod r and the onsite energy is linear$\omega_0=ar+b$
It is worth mentioning that the qualitative nature of the defect modes can be tuned by adjusting the defect rod radius. For different range of radii, it can admit $s,p,d$ type orbitals\cite{villeneuve1996microcavities}. For simplicity, in this work we shall only consider $s$-type of defect modes arising from rods with radius $r<0.2a$. The defect band depends on the radius of the defect rod. Fig. \ref{fig:1}b shows that the frequency of the defect mode is a monotonic function of the radius of the defect rod.

\section{Defect modes of the Hofstadter model protected by a factorizable second Chern number with abelian Berry curvature}
\begin{figure}[H]
\centering
\includegraphics[width=\linewidth]{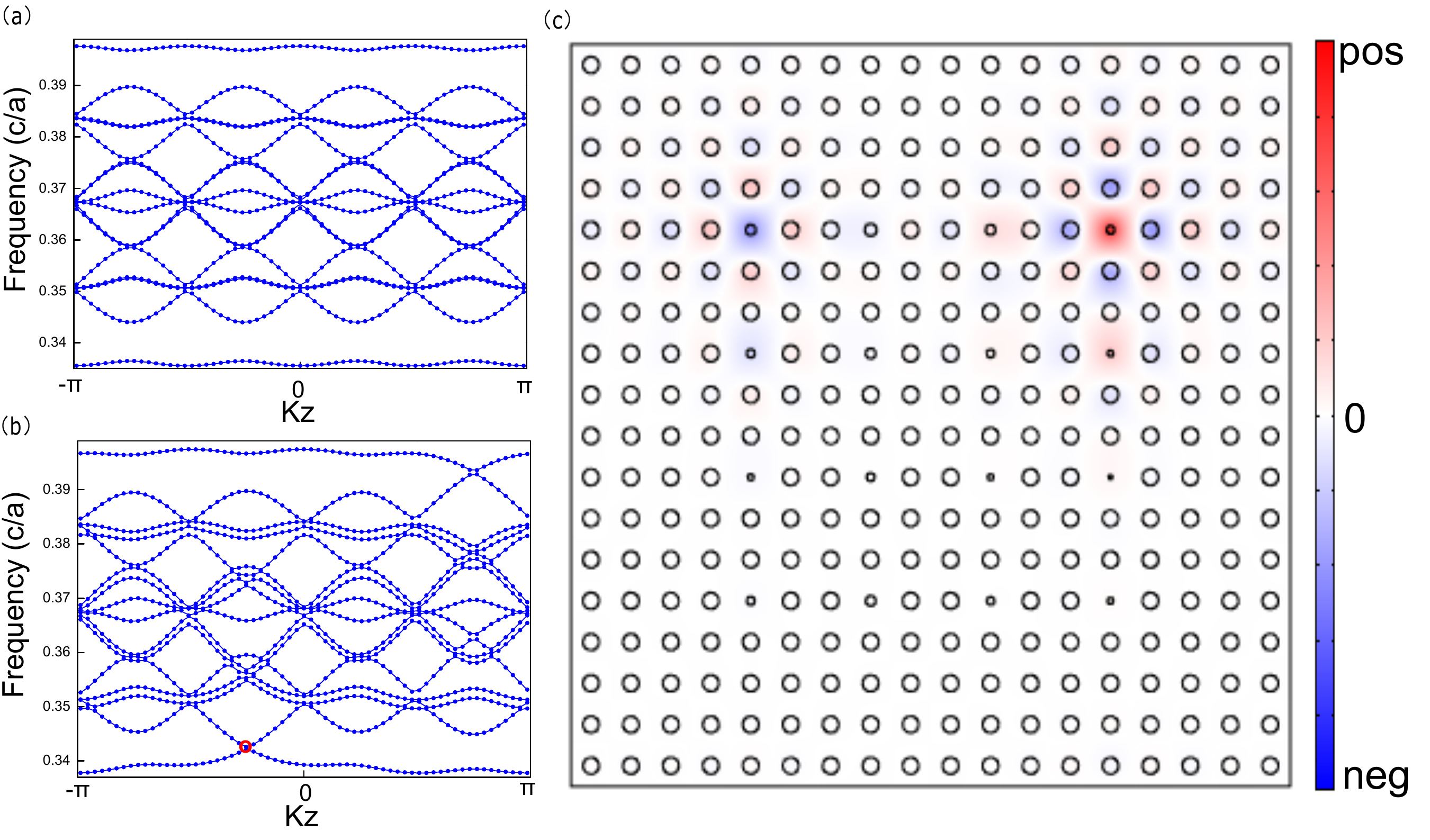}
\caption{(a) and (b): The frequency bands of the defect states of our 2d modulated PC resonator lattice embedded in a regular 2d PC, computed with periodic and open boundary conditions respectively. They are plotted as a function of modulation offset parameter $k_z$, with $k_w$,$k_x$,$k_y$ set to $0$. The radius of the defect rods obeys Eq.\ref{eq3} with parameters $r_{0}=0.1a$, $r_{1}=r_{2}=0.025a$ and $b=1/4$. For a), The Maxwell's equation is solved with the FEM is performed with a repeated (periodic boundary condition) super-cell having 16 defect rods with $n=4$ along $x$ and $y$. By fitting the band data with Eq.\ref{TB2}. We found that $\omega_0=0.36862(c/a)$ and $\lambda=-0.014508$ and $t=-0.00398(c/a)$. (c) The $z$ direction electric field strength ($E_z$) distribution of the defect boundary state at $k_z=-0.8$ and $\omega=0.3423(c/a)$ (red circle in (b)) with open boundary conditions. The defect states are evidently localized at the edge in both $x$ and $y$ directions, with the directional asymmetry in decay length arising from the inequivalence of $k_z$ and $k_w$. The maximum value in the color bar is 20V/m while the minimum value in the color bar is -20V/m.}
\label{fig:3}
\end{figure}

To realize PC defect states that are protected by the factorizable second Chern number, we simply modulate the size of the resonators such that we obtain an effective Hofstadter model\cite{kraus2012topological,madsen2013topological}. Here we illustrate the simple example of 2d defect modes protected by the second Chern number shown in Fig. \ref{fig:3}. We construct this defect lattice using a similar setup as Fig. \ref{fig:1}, but modulate the radius of the resonators instead of keeping them constant:
\begin{equation}
r=r_{0}+r_{1} \cos(2\pi b x+k_z) +r_{2} \cos(2\pi b y+k_w),
\label{eq3}
\end{equation}
where the tuple $x$ and $y$ describes a resonator located on the $3x^{th}$ row and $3y^{th}$ column. $b$ is a rational number which we keep as $b=p/q=1/4$. The synthetic dimension parameters $k_z$ and $k_w$ sets the radius for the first defect, which is given by Eq.\ref{eq3}. The defect states are numerically solved with the FEM method with an $q \times q$ resonator 2d super-cell. In Fig. \ref{fig:3}a, when we apply a periodic boundary condition for the super-cell, which means we repeat the super-cell structure in $x$ and $y$ direction, we can see two band gaps in the 2d Brillounin zone (BZ) of $k_z$ and $k_w$ (Fig. \ref{fig:3}a). However, if we take an open boundary condition in $x$ and $y$ direction, which means the surrounding of the super-cell is terminated with regular dielectric rods constituting the bulk (Fig. \ref{fig:3}c). Then we see the boundary states appearing in the gap (Fig. \ref{fig:3}b), whose field distribution is shown in Fig. \ref{fig:3}c.  From Fig. \ref{fig:3}, the bandstructure (for $b=1/4$) consists of a fractal hierarchy of bulk bands. Upon varying $k_z$ over a period, edge states are seen to continuously connect the bulk bands. 
To understand the nature of the defect edge modes, we write down the effective Hamiltonian which should be of the form
\begin{equation}
\begin{split}
H&=\sum_{x,y}(\omega_0+\lambda \cos(2\pi bx+k_z))\left|\psi_{(x,y)}\right\rangle \left\langle \psi_{(x,y)}\right|\\
&+\sum_{x,y}(\lambda \cos(2\pi by+k_w))\left|\psi_{(x,y)}\right\rangle \left\langle \psi_{(x,y)}\right|\\
&+\sum_{x,y} t \left|\psi_{(x,y)}\right\rangle \left\langle \psi_{(x+ 1,y)}\right|+\sum_{x,y} t \left|\psi_{(x,y)}\right\rangle \left\langle \psi_{(x,y+ 1)}\right| \\
&+\sum_{y}t e^{i q k_x}\left|\psi_{(q,y)}\right\rangle \left\langle \psi_{(1,y)}\right|+\sum_{x}t e^{i q k_y}\left|\psi_{(x,q)}\right\rangle \left\langle \psi_{(x,1)}\right|\\
&+h.c.
\end{split}
\label{TB2}
\end{equation}
which follows from the form of the periodic modulation in the defect radius (Eq. \ref{eq3}). In the s-wave approximation, the hopping between nearest resonators are constant $t$ to leading order, and further hoppings can be dropped. This is just a superposition of two (off-diagonal) Aubry-Andr\'e models\cite{AAH} in different directions, with eigenmodes being products of the 1d Aubry-Andr\'e model eigenmodes, which is also constructed in the Supplementary Material.
Upon reinterpreting the basis modes in Eq. \ref{TB2} as Laudau gauge eigenfunctions, our 2d defect PC lattice can be interpreted as a tensor product of two Hofstadter models, which hence lives in 4d space. This allows for a natural interpretation of its topology as arising from the second Chern number\cite{kraus2013four,qi2008topological} 
\begin{equation}
C_2=\frac{\epsilon_{ijkl}}{32\pi^2}\int Tr[f_{ij}f_{kl}]~d^4k=\frac{1}{4\pi^2}\int f_{xz}~d^2k\int f_{yw}~d^2k
\end{equation}
where $f_{ij}=\partial_iA_{j}-\partial_jA_{i}$  is the abelian berry curvature for $i,j$ dimensions of the Hofstadter model, and $A_{i}=\langle \psi(\vec{k})|\partial_i|\psi(\vec{k})\rangle$ is the abelian berry connection. In our case, $C_2$ is particularly easy to compute by Fukui’s method, since the product nature of the eigenmodes implies that $Tr(f\wedge f)$ factorizes into two copies of $Tr f$. So the second Chern number $C_2$ can be written as the product of two first Chern number$\int f_{xz}~d^2k$ and $\int f_{yw}~d^2k$. The detailed calculation of both first and second Chern number using Fukui's method is shown in the Supplementary Material, in which the second Chern number is given by
\begin{equation}
\nu=\nu_{1}+\nu_{2}+\nu_{3}=\sum_{\vec{k}}\mathcal{F}_{zx}\mathcal{F}_{yw}+\mathcal{F}_{yz}\mathcal{F}_{xw}+\mathcal{F}_{xy}\mathcal{F}_{zw}
\end{equation}
The numerical result indeed proves $\nu_1=-1$, $\nu_2=0$ and $\nu_3=0$.

\section{Defect modes protected by a non-factorizable second Chern number with non-abelian Berry curvature}
\begin{figure}[H]
\centering
\includegraphics[width=\linewidth]{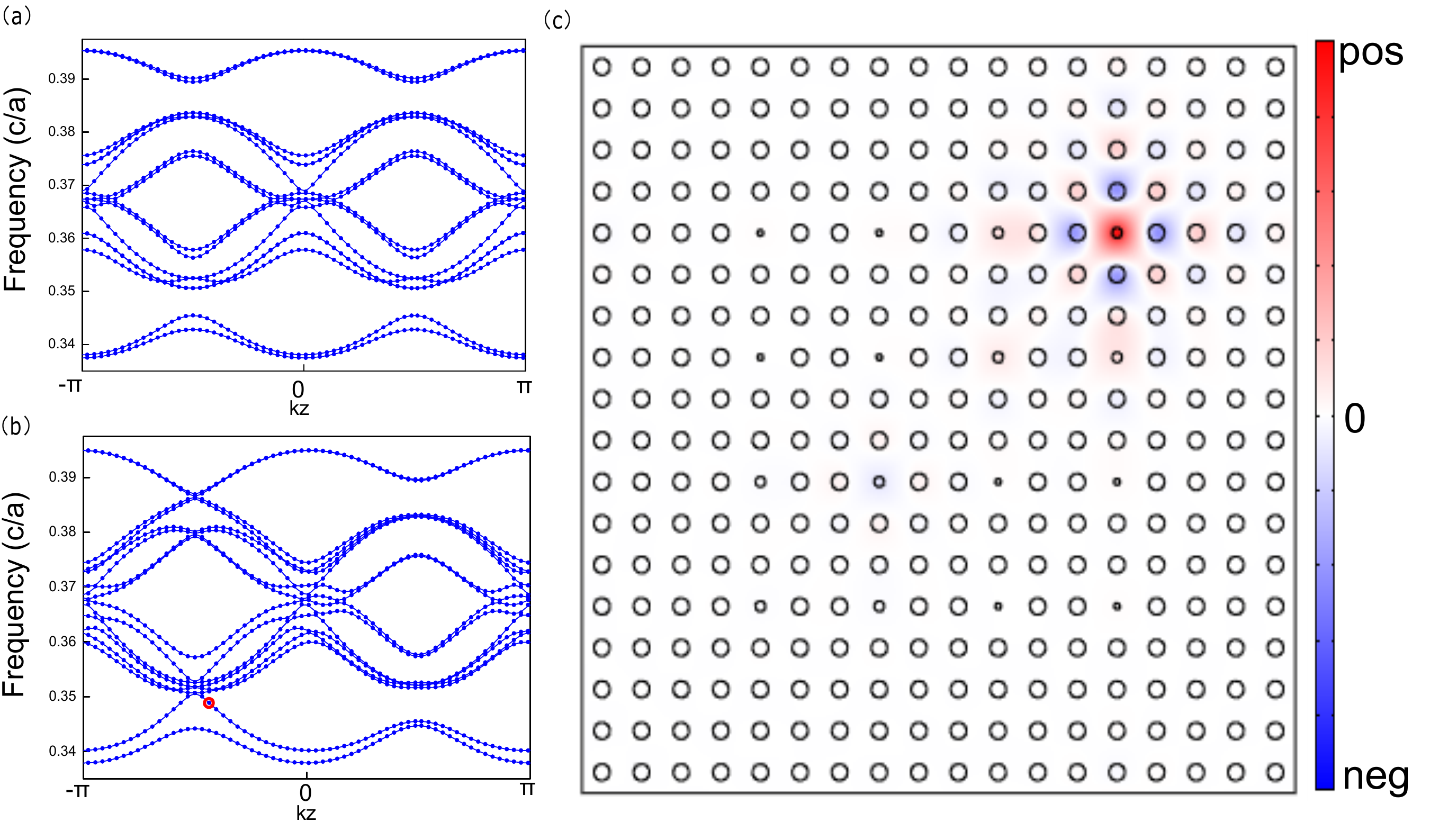}

\caption{(a) and (b): The frequency bands of the defect states of our 2d modulated PC resonator lattice embedded in a regular 2d PC, computed with periodic and open boundary conditions respectively. They are plotted as a function of modulation offset parameter $k_z$, with $k_w$,$k_x$,$k_y$ set to $0$. The radius of the defect rods obeys Eq.\ref{eq4} with parameters $r_{0}=0.1a$, $r_{1}=r_{2}=0.025a$ and $b=1/4$. For a), The Maxwell's equation is solved with the FEM is performed with a repeated (periodic boundary condition) super-cell having 64 defect rods with $n=8$ along $x$ and $y$. By fitting the band data with eq1. We found that $\omega_0=0.36862(c/a)$ and $\lambda=-0.014508$ and $t=-0.00398(c/a)$. (c) The $z$ direction electric field strength ($E_z$) distribution of the defect boundary state at $k_z=-1.37$ and $\omega=0.3671(c/a)$ (red circle in (b)) with open boundary conditions. The defect states are evidently localized at the edge in both $x$ and $y$ directions. The maximum value in the color bar is 20V/m while the minimum value in the color bar is -20V/m}
\label{fig:4}
\end{figure}

We now come to the main results of this paper, which involves direct generalization of the factorizable second Chern number defect lattice construction, through modulating the radius of the defect rods shown in Fig. \ref{fig:4}:
\begin{equation}
r=r_{0}+r_{1} \cos(2\pi b (x+y)+k_z) +r_{2} \cos(2\pi b (x-y)+k_w),
\label{eq4}
\end{equation}
where the tuple $x$ and $y$ describes a resonator located on the $3x^{th}$ row and $3y^{th}$ column. $b$ is a rational number which we keep as $b=p/q=1/4$. The synthetic dimension parameters $k_z$ and $k_w$ sets the radius for the first defect, which is given by Eq.\ref{eq4}. The defect states are numerically solved with the FEM method with an $q \times q$ resonator 2d super-cell. In Fig. \ref{fig:4}a, when we apply a periodic boundary condition for the super-cell, which means we repeat the super-cell structure in $x$ and $y$ direction, we can see two band gaps in the 2d BZ of $k_z$ and $k_w$ (Fig. \ref{fig:4}a). However, if we take an open boundary condition in $x$ and $y$ direction, which means the surrounding of the super-cell is terminated with regular dielectric rods constituting the bulk (Fig. \ref{fig:4}c). Then we see the boundary states appearing in the gap (Fig. \ref{fig:4}b), whose field distribution is shown in Fig. \ref{fig:4}c.
To understand the nature of the defect edge modes, we write down the effective Hamiltonian which should be of the form
\begin{equation}
\begin{split}
H&=\sum_{x,y}(\omega_0+\lambda( \cos(2\pi b(x+y)+k_z))\left|\psi_{(x,y)}\right\rangle \left\langle \psi_{(x,y)}\right|\\
&+\sum_{x,y}( \cos(2\pi b(x-y)+k_w))\left|\psi_{(x,y)}\right\rangle \left\langle \psi_{(x,y)}\right|\\
&+\sum_{x,y} t \left|\psi_{(x,y)}\right\rangle \left\langle \psi_{(x+ 1,y)}\right|+\sum_{x,y} t \left|\psi_{(x,y)}\right\rangle \left\langle \psi_{(x,y+ 1)}\right| \\
&+\sum_{y}t e^{i q k_x}\left|\psi_{(q,y)}\right\rangle \left\langle \psi_{(1,y)}\right|+\sum_{x}t e^{i q k_x}\left|\psi_{(x,q)}\right\rangle \left\langle \psi_{(x,1)}\right|\\
&+h.c.
\end{split}
\label{TB2}
\end{equation}
which follows from the form of the periodic modulation in the defect radius (Eq. \ref{eq4}). In the s-wave approximation, the hopping between nearest resonators are constant $t$ to leading order, and further hoppings can be dropped. Different from the Hofstadter model, its edge state is protected by a non-factorizable second Chern number with non-abelian Berry curvature
\begin{equation}
C_2=\frac{\epsilon_{ijkl}}{32\pi^2}\int Tr[f^{\alpha \beta}_{ij}f^{\alpha \beta}_{kl}]~d^4k
\end{equation}
where $f_{ij}^{\alpha \beta}=\partial_iA_{j}^{\alpha \beta}-\partial_jA_{i}^{\alpha \beta}+i[A_{i},A_{j}]^{\alpha \beta}$  is the non-abelian berry curvature for $i,j$ dimensions of the model. It is defined by the Berry connection  $A_{i}^{\alpha \beta}=\langle \psi^{\alpha}(\vec{k})|\partial_i|\psi^{\beta}(\vec{k})\rangle$of the two bands below the gap. By computing $C_2$ by Fukui’s method (see Supplementary Material for details), we find that $Tr(f\wedge f)$ cannot factorizes into two copies of $Tr f$. In fact the second Chern number is given by
\begin{equation}
\nu=\nu_{1}+\nu_{2}+\nu_{3}=\sum_{\vec{k}}\mathcal{F}_{zx}\mathcal{F}_{yw}+\mathcal{F}_{yz}\mathcal{F}_{xw}+\mathcal{F}_{xy}\mathcal{F}_{zw}
\end{equation}
 and the numerical result shows $\nu_1=1$ and $\nu_2=1$ and $\nu_3=0$.

\section{experimental proposal}

Here we describe a simple experimental proposal for realizing these edge states through transmission spectrum measurements in the microwave wavelength window\cite{Maser1,Maser2,Maser3}.

Our PC is made up of a background ``vacuum'' 2d lattice of rods of dielectric constant $8$ with lattice constant $a=0.104$ m and rod radius $0.2a$. One in every three rods is replaced by a defect rod of the same dielectric constant, but of radius periodically modulated as per Eq. \ref{eq3} and Eq \ref{eq4} and Fig. \ref{fig:3} and Fig. \ref{fig:4}. This PC can be easily constructed with conventional PC fabrication processes. It is very easy to fabricate the PC sample with high precision since $a$ is in the decimeter range.

A midgap edge state will appear as a distinct peak in the transmission spectrum when the wavelength is tuned within the bulk gap. From our FEM simulations we have, for instance, in the Hofstadter model case the resonant peak $\lambda=0.3038 m$ when $k_z=-0.8,k_w=0$. We can observe that topological edge states appear within the bulk gaps due to the presence of an effective magnetic flux $b$ which takes the form of defect modulation. Similarly for our new model with non-factorizable second Chern number, the resonant peak $\lambda=0.2833 m$ when $k_z=-1.37,k_w=0$.

\section{Conclusion}
In summary, we have proposed a way to realize both factorizable and non-factorizable 2nd Chern number topological phases with regular, easily-fabricated photonic crystals with defect resonators (rods) embedded within a regular lattice of resonators. Through quasiperiodic spatial modulations in the defect radii, in the former case the defect lattice can be described by a direct product of effective Hofstadter models possessing topologically nontrivial Chern bands with abelian Berry curvature. In the latter case, we realize a new model with nontrivial Chern bands of non-abelian Berry Curvature. The topological signatures of these bands lie in their edge states, which are seen to traverse the bulk gaps as synthetic dimension parameters $k_z,k_w$ are varied over a period. Such photonic systems can be easily experimentally realized with regular photonic crystals with dielectric rods in air, and we propose experiments realizing our precise predictions from FEM simulations. Those topologically protected defect modes has potential applications  due to their immunity to small perturbations.

\section{ACKNOWLEGMENT}
Bochen Guan is supported by a scholarship from the Oversea Study Program of Guangzhou Elite Project.
\section{Supplymentary Material}
\section{Defect modes protected by the first Chern number}

\begin{figure}[H]
\centering
\includegraphics[width=\linewidth]{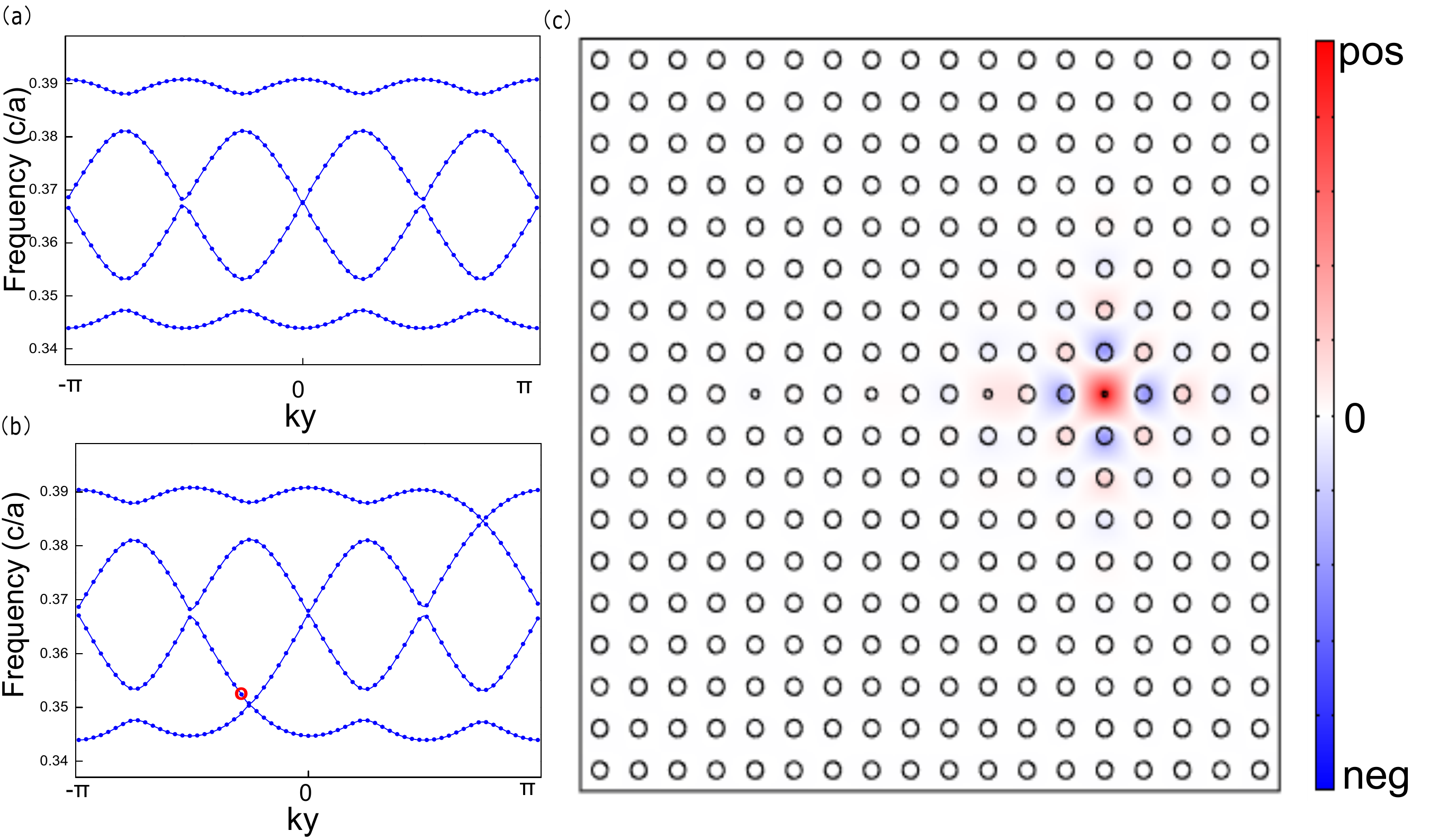}
\caption{ (a) The frequency bandstructure of the defect states for a 1d PC resonator lattice embedded in a regular periodic 2d PC, as a function of $k_y$($k_x=0$). The radius of the defect rods obeys Eq.\ref{eq1} with parameters $r_0=0.1a$, $r_{1}=0.03a$ and $b=p/q=1/4$. The FEM is performed with a repeated (periodic boundary condition) super-cell having $q=4$ defect rods. By fitting the band structure with tight binding model we have $\omega_{0}=0.36862(c/a)$ and $\lambda=-0.0174096$. (b) The frequency bands of defect states for a 1d PC resonator lattice with open boundary condition as a function of $k_y $($k_x=0$). There is only one super-cell terminated by regular bulk dielectric rods on both ends. (c) The $z$ direction electric field strength ($E_z$) distribution of the localized defect boundary state at $k_y=1.1$ and $\omega=0.3480 (c/a)$ (red circle in (b)) with open boundary conditions. }
\label{fig:2}
\end{figure}

To realize PC defect states that are protected by the first Chern number, we simply modulate the size of the resonators such that we obtain an effective Hofstadter model\cite{kraus2012topological,madsen2013topological}. Here we illustrate the simple example of 1d defect modes protected by the first Chern number shown in (Fig. \ref{fig:2}). We construct this defect lattice using a similar setup as Fig. \ref{fig:1}, but modulate the radius of the resonators instead of keeping them constant:
\begin{equation}
r=r_{0}+r_{1} \cos(2\pi b x+k_y),
\label{eq1}
\end{equation}
where $x$ labels the resonators in the $\hat x$ direction and $b$ is an rational number. A synthetic dimension is implemented through the parameter $k_y$, which sets the radius of the first defect. From Fig. \ref{fig:2}, the bandstructure (for $b=1/4$) consists of a fractal hierarchy of bulk bands. Upon varying $k_y$ over a period, edge states are seen to continuously connect the bulk bands. To understand the topological origin of these mid-gap states, one refers to the effective TB Hamiltonian 
\begin{equation}
\begin{split}
H&=\sum_{x}\lambda\cos(2\pi b x + k_y)|\psi_{x}\rangle \langle \psi_{x}|\\
&+\sum_{x,x+1}t |\psi_{x}\rangle \langle \psi_{x+1}|\\
&+t e^{-i q k_x}|\psi_{x}\rangle \langle \psi_{1}|+h.c
\end{split}
\label{AA}
\end{equation}
which follows from the form of the periodic modulation in the defect radius (Eq.\ref{eq1}). In the s-wave approximation, the hopping between nearest resonators are constant $t$ to leading order, and further hoppings can be dropped. Hence we recover a variation of the 1d Aubry-Andr\'e (AA) model\cite{AAH}. Eq.\ref{AA} can be mapped to a 2d integer quantum Hall system by reinterpreting the modes $|\psi_{x}\rangle$ as Landau gauge wavefunctions on a 2d lattice. If we also reinterpret $k_y$ as a momentum variable $k_ya$, Eq. \ref{AA} becomes the Hofstadter Hamiltonian. This is just the hopping Hamiltonian on a rectangular lattice with horizontal and vertical hoppings $t$ and $\lambda/2$, and $2\pi b$ flux threaded per plaquette, i.e. a quantum Hall system on a lattice. When the flux $b=p/q$ is rational where $p$ and $q$ are relatively prime, this hopping Hamiltonian describes a Chern insulator with $q$ bands, each with Chern numbers detailed as in Ref. \onlinecite{hatsugai1993edge}. If $b$ is irrational, we can only approximately say $b\approx p/q$. In the irrational limit, $q$ diverges and the bands tend towards perfectly flat 'Landau levels'\footnote{Indeed, an infinite number of bands is required for perfectly flat Chern bands, as can be proven via K-theory\cite{chen2014impossibility,lee2016band,read2016compactly}. }.

Viewed as a Chern insulator, the midgap states are just the topological edge states corresponding to the bulk chiral anomalies\cite{qi2008topological}. Physically, they realize the pumping mechanism in Laughlin's spectral flow argument, even though in this context it is the defect modes and not electrons that are pumped. These midgap states exist at the edge (ends) of the defect line, as is evident in Fig. \ref{fig:2}c), and are pumped from one end to the other as the midgap state traverse a gap\cite{huang2012entanglement,alexandradinata2014wilson,lee2015free}. The number of states pumped over a period of $k_y$ within a gap depends on the difference in the combined Chern numbers of the bands on either side of gap. 

For our FEM solution to Maxwell's equations, we take $q=4$ resonators in a 1d super-cell.
As illustrated in Fig. \ref{fig:2}a, when we apply a periodic boundary condition for the super-cell, which means we repeat the super-cell structure in $x$ direction, we can see bulk bands separated by band gaps in the 1d BZ of $k_y$ (Fig. \ref{fig:2}a). However, if we take an open boundary condition in the $x$ direction, which means the that left and right sides of the super-cell are terminated with regular dielectric rods in the bulk (Fig. \ref{fig:2}c), we see the boundary states (Fig. \ref{fig:2}b) localized on the end of defect chain, whose field distribution is shown in Fig. \ref{fig:2}c.

\section{Numerical Method for calculating the first Chern Number}
We use The Fukui’s method to compute first Chern Number\cite{fukui2005chern,PhysRevLett.115.195303}. The  BZ in this model is $[0,\frac{2 \pi}{q}]\times [0,\pi] $ and discretized as $k_{\mu}=\frac{2 \pi N}{q N_b}$ with $\mu=x$ and $N=0,1,…..,N_b -1$ and $k_{\nu}=\frac{2 \pi N}{q N_b}$ with $\nu=y$ and $N=0,1,…..,N_b q-1$ (we take Nb=50). The  absolute  value  of displacement  in $\vec{e}_x$,$\vec{e}_y$ direction is $\epsilon=\frac{2 \pi}{q N_b}$. We define the link variable
\begin{equation}
U_{\mu}=\frac{\langle u(\vec{k})|u(\vec{k}+\epsilon \vec{e}_{\mu})\rangle}{|\langle u(\vec{k})|u(\vec{k}+\epsilon \vec{e}_{\mu})\rangle |}
\end{equation}
$|u(k)\rangle$ is the wavefunction,  a $q$ component vector. The Berry curvature can be written as 
\begin{equation}
\mathcal{F}_{\mu \nu}=\frac{1}{2 \pi i} log\frac{U_{\mu}(\vec{k}) U_{\nu}(\vec{k}+\epsilon \vec{e}_{\mu})}{ U_{\mu}(\vec{k}+\epsilon \vec{e}_{\nu}) U_{\nu}(\vec{k})}
\end{equation}
The First Chern number is given by
\begin{equation}
\nu=\sum_{\vec{k}} \mathcal{F}_{xy}
\end{equation}
The numerical result gives $\nu_1=1$.

\section{method for computing the factorizable second Chern Number with abelian Berry curvature}
\begin{figure}[H]
\centering
\includegraphics[width=\linewidth]{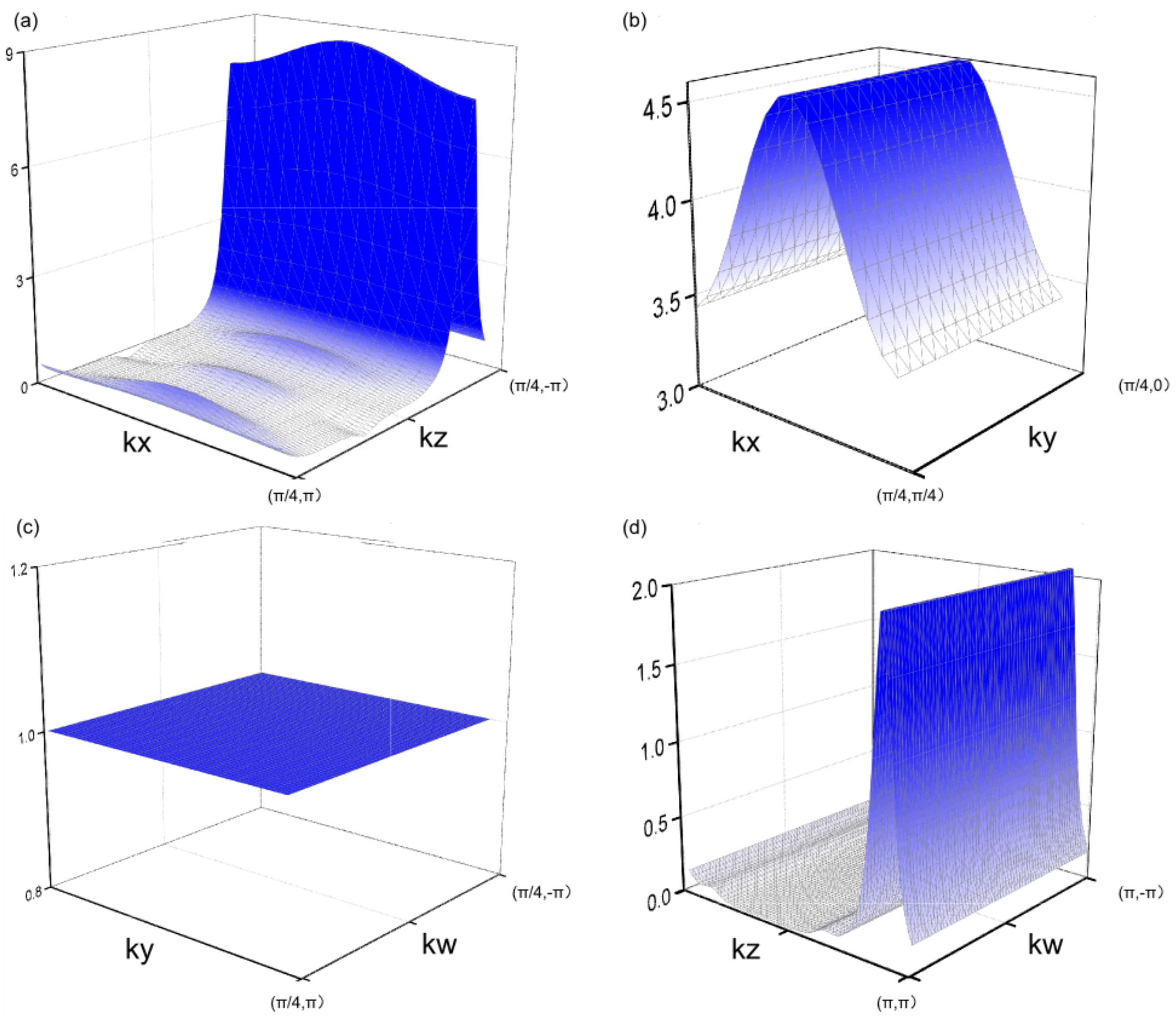}
\caption{Since all the Berry curvatures functions of four dimensional variables, so we can only plot them by integrating out two dimensions. Here we plot integrated forms of $f_{xz}$ as an example. (a)The function  of $\int f_{xz} dk_{y} dk_{w}$. (b)The function  of $\int f_{xz} dk_{z} dk_{w}$. (c)The function  of $\int f_{xz} dk_{x} dk_{z}$. (d)The function  of $\int f_{xz} dk_{x} dk_{y}$. }
\label{fig:productive}
\end{figure}

We use the Fukui’s method to compute the second Chern Number\cite{fukui2005chern,PhysRevLett.115.195303}. The  BZ in this model is $[0,\frac{2 \pi}{q}]\times[0,\frac{2 \pi}{q}]\times [0,2\pi]\times [0,\pi] $ and discretized as $k_{\mu}=\frac{2 \pi N}{q N_b}$ with $\mu=x,y$ and $N=0,1,…..,N_b -1$ and $k_{\nu}=\frac{2 \pi N}{q N_b}$ with $\nu=z,w$ and $N=0,1,…..,N_b q-1$  (Nb=16). The absolute value of the displacement in $\vec{e}_x$,$\vec{e}_y$,$\vec{e}_z$,$\vec{e}_w$ direction is $\epsilon=\frac{2 \pi}{q N_b}$. We define the link variable
\begin{equation}
U_{\mu}=\frac{\langle u(\vec{k})|u(\vec{k}+\epsilon \vec{e}_{\mu})\rangle}{|\langle u(\vec{k})|u(\vec{k}+\epsilon \vec{e}_{\mu})\rangle |}
\end{equation}
$|u(k)\rangle$ is the wavefunction with $q^2$ components. The abelian Berry curvature can be written as 
\begin{equation}
\mathcal{F}_{\mu \nu}=\frac{1}{2 \pi i} log\frac{U_{\mu}(\vec{k}) U_{\nu}(\vec{k}+\epsilon \vec{e}_{\mu})}{ U_{\mu}(\vec{k}+\epsilon \vec{e}_{\nu}) U_{\nu}(\vec{k})}
\end{equation}
The Second Chern number is given by
\begin{equation}
\nu=\nu_{1}+\nu_{2}+\nu_{3}=\sum_{\vec{k}}\mathcal{F}_{zx}\mathcal{F}_{yw}+\mathcal{F}_{yz}\mathcal{F}_{xw}+\mathcal{F}_{xy}\mathcal{F}_{zw}
\end{equation}
The numerical result gives $\nu_1=-1$ and $\nu_2=0$ and $\nu_3=0$. To visualize the calculation, we plot one Berry curvatures in an integrated form in Fig. \ref{fig:productive} as an example.

\section{method for computing non-factorizable second Chern Number with non-abelian curvature}
\begin{figure}[H]
\centering
\includegraphics[width=\linewidth]{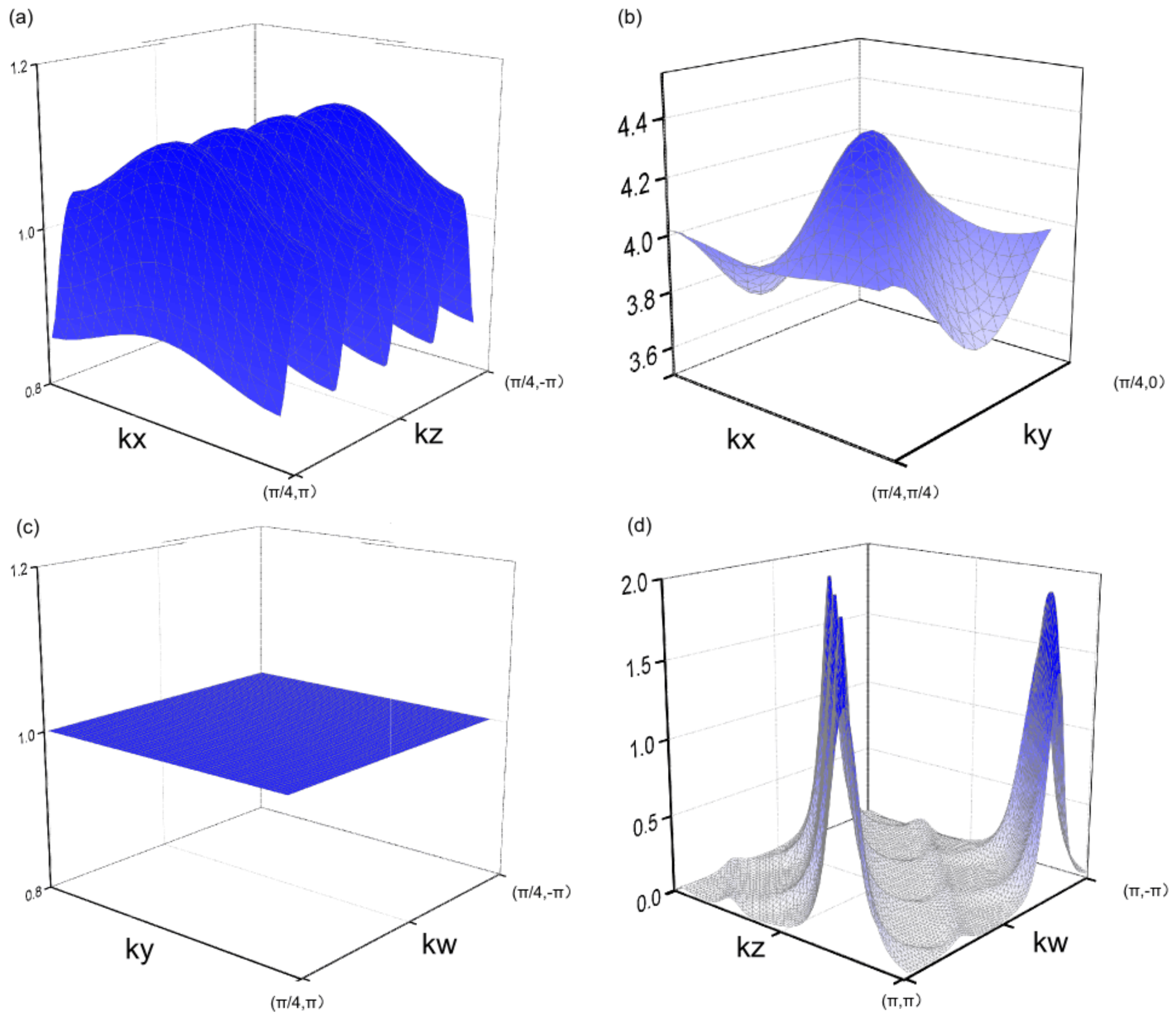}
\caption{Since all the Berry curvatures functions of four dimensional variables, so we can only plot them by integrating out two dimensions. Here we plot integrated forms of $f_{xz}$ as an example. (a)The function  of $\int f_{xz} dk_{y} dk_{w}$. (b)The function  of $\int f_{xz} dk_{z} dk_{w}$. (c)The function  of $\int f_{xz} dk_{x} dk_{z}$. (d)The function  of $\int f_{xz} dk_{x} dk_{y}$. }
\label{fig:nonproductive}
\end{figure}

We have two bands below the lowest band gap which correspond to states $\psi_{a}(\vec{k})$ and $\psi_{b}(\vec{k})$. Similar to the abelian case we can discretize the BZ as $k_{\mu}=\frac{2 \pi N}{q N_b}$ with $\mu=x,y$ and $N=0,1,…..,N_b -1$ and $k_{\nu}=\frac{2 \pi N}{q N_b}$ with $\nu=z,w$ and $N=0,1,…..,N_b q-1$ (Nb=16). The absolute value of displacement in $\vec{e}_x$,$\vec{e}_y$,$\vec{e}_z$,$\vec{e}_w$ direction is $\epsilon=\frac{2 \pi}{q N_b}$. We define the link variable\cite{fukui2005chern}
\begin{equation}
U_{\mu}=\frac{\begin{bmatrix}
\langle \psi_{a} (\vec{k})|\psi_{a} (\vec{k}+\epsilon \vec{e}_{\mu}) \rangle &\langle \psi_{a} (\vec{k})|\psi_{b} (\vec{k}+\epsilon \vec{e}_{\mu}) \rangle \\
\langle \psi_{b} (\vec{k})|\psi_{a} (\vec{k}+\epsilon \vec{e}_{\mu}) \rangle & \langle \psi_{b} (\vec{k})|\psi_{b} (\vec{k}+\epsilon \vec{e}_{\mu}) \rangle \\	
\end{bmatrix} }{det\begin{bmatrix}
\langle \psi_{a} (\vec{k})|\psi_{a} (\vec{k}+\epsilon \vec{e}_{\mu}) \rangle &\langle \psi_{a} (\vec{k})|\psi_{b} (\vec{k}+\epsilon \vec{e}_{\mu}) \rangle \\
\langle \psi_{b} (\vec{k})|\psi_{a} (\vec{k}+\epsilon \vec{e}_{\mu}) \rangle & \langle \psi_{b} (\vec{k})|\psi_{b} (\vec{k}+\epsilon \vec{e}_{\mu}) \rangle \\	
\end{bmatrix} }
\end{equation}
$|\psi_{a}(k)\rangle$ and $|\psi_{b}(k)\rangle$ are  wavefunctions with $q^2$ components each. The abelian Berry curvature can be written as 
The non-abelian Berry curvature can be written as
\begin{equation}
\mathcal{F}_{\mu \nu}=\frac{1}{2 \pi i} log\frac{U_{\mu}(\vec{k}) U_{\nu}(\vec{k}+\epsilon \vec{e}_{\mu})}{ U_{\mu}(\vec{k}+\epsilon \vec{e}_{\nu}) U_{\nu}(\vec{k})}
\end{equation}
The second Chern number is given by
\begin{equation}
\nu=\nu_{1}+\nu_{2}+\nu_{3}=\sum_{\vec{k}}\mathcal{F}_{zx}\mathcal{F}_{yw}+\mathcal{F}_{yz}\mathcal{F}_{xw}+\mathcal{F}_{xy}\mathcal{F}_{zw}
\end{equation}
The numerical result gives $\nu_1=1.0005\approx 1$ and $\nu_2=1.0005\approx 1$ and $\nu_3=0$. To visualize the calculation, we plot one Berry curvatures in an integrated form in Fig. \ref{fig:nonproductive} as an example.
To calculate the non-abelian Berry curvature, we consider two bands below the band with mini-gap between them. In fact, if we calculate their Chern number separately by Fukui's method in the abelian way, it will not converge to an integer even if Nb goes to infinity.
\bibliography{apssamp}
\end{document}